\documentclass[conference]{IEEEtran}
\usepackage{amsmath,amssymb,amsfonts}
\usepackage{algorithmic}
\usepackage{graphicx}
\usepackage{hyperref}
\usepackage{textcomp}
\usepackage{multicol}
\usepackage{float}
\usepackage{url}
\usepackage{pgfplots} 
\usepackage{xcolor}
\usepackage{pgf-pie}
\usetikzlibrary{patterns}
\usepackage{tikz}

\AtBeginDocument{%
  \providecommand\BibTeX{{%
    \normalfont B\kern-0.5em{\scshape i\kern-0.25em b}\kern-0.8em\TeX}}}

    \makeatletter
\newcommand{\linebreakand}{%
  \end{@IEEEauthorhalign}
  \hfill\mbox{}\par
  \mbox{}\hfill\begin{@IEEEauthorhalign}
}

\makeatother
\IEEEoverridecommandlockouts

\begin{document}

\title{Software Companies' Responses to Hybrid Working*}

\author{\IEEEauthorblockN{Dron Khanna\thanks{*This is the authors’ version of the manuscript accepted for publication in the Proceedings of 50th Euromicro Conference Series on Software Engineering and Advanced Applications (SEAA) 2024. This manuscript version is made available under the CC-BY-NC-ND4.0 license. http://creativecommons.org/ licenses/by-nc-nd/4.0}}
\IEEEauthorblockA{\textit{Free University of Bozen-Bolzano}\\
Bolzano, Italy \\
dron.khanna@unibz.it}
\and
\IEEEauthorblockN{Henry Edison}
\IEEEauthorblockA{
\textit{Blekinge Institute of Technology}\\
Karlskrona, Sweden \\
henry.edison@bth.se}
\linebreakand
\IEEEauthorblockN{Anh Nguyen-Duc}
\IEEEauthorblockA{\textit{University of South Eastern Norway}\\
Bo i Telemark, Norway \\
Anh.Nguyen.duc@usn.no}
\and 
\IEEEauthorblockN{Kai-Kristian Kemell}
\IEEEauthorblockA{\textit{University of Helsinki}\\
Helsinki, Finland \\
kai-kristian.kemell@helsinki.fi} }

\maketitle
\begin{abstract}
[Context:] COVID-19 pandemic has disrupted the global market and workplace landscape. As a response, hybrid work situations have become popular in the software business sector. This way of working has an impact on software companies. [Objective]: This study investigates software companies' responses to hybrid working. [Method]: We conducted a large-scale survey to achieve our objective. Our results are based on a qualitative analysis of 124 valid responses. [Results]: The main result of our study is a taxonomy of software companies’ impacts on hybrid working at individual, team and organisation levels. We found higher positive responses at individual and organisational levels than negative responses. At the team level, both positive and negative impacts obtained a uniform number of responses. [Conclusion]: The results indicate that hybrid working became credible with the wave of COVID-19, with 83 positive responses outweighing the 41 negative responses. Software company respondents witnessed better work-life balance, productivity, and efficiency in hybrid working.
\end{abstract}

\begin{IEEEkeywords}
flexible work, hybrid work, remote work, work from home (WFH), online work
\end{IEEEkeywords}


\section{Introduction}
For many, the office was a second home in the pre-pandemic period. Bustling and towering high-rise offices were filled with several employees till 2019 \footnote{https://www.fdiintelligence.com/content/data-trends/office-imbalance-us-vacancy-rates-hit-record-high-82348}. Early 2020, the start of the COVID-19 pandemic, the office routine faded and hybrid work routines such as remote work, work from home (WFH) or hybrid work \cite{kemell2023hybrid}, \cite{smite199} became the norm wherever it was acceptable, such as in the Software Engineering (SE) field \cite{russo_21}. Currently, with various perceptions of hybrid work \cite{atiku2020perceptions}, the work arrangement continues to be expected \cite{smite_new} as employees prefer more freedom to either work from home or the office with fewer restrictions and better opportunities to socialize with colleagues \cite{nolan}. Though it was the pandemic that brought a change in the hybrid work\cite{kemell2023hybrid},\cite{smite199},\cite{mcphail2024post} (also the term is often used and interpreted in different ways, for example, flexible work, remote work, WFH \cite{molino}, telecommuting \cite{allen}, and telework \cite{ruth}, \cite{smite199}) pattern into the spotlight, is not at all a new phenomenon. Thousands of teleworkers have previously worked remotely and in multinational software organisations with various practices and tools dealing with geographically distributed teams \cite{bailey}.

However, previous pandemic studies have focused mainly on the productivity of individual employees (e.g., \cite{west}, \cite{butler}, \cite{bloom}, \cite{edison2015towards}). The COVID-19 pandemic aftermath led the research interest across various disciplines. Therefore, some studies have explored both developers' well-being and individual developers' productivity in the COVID-19 Work-From-Home context (e.g., \cite{ralph_pp},\cite{canna}). Whereas few also focused on the satisfaction and performance of developers \cite{russo_223} and specifically on the work practices related to remote work (e.g., \cite{argen}, \cite{miller},\cite{bez_sdt}). Regarding positive or negative effects of the pandemic way of working, the study by Canna et al. listed financial and job security concerns as one negative effect concerning developers' well-being \cite{canna}.

As we enter the post-pandemic era \cite{mcphail2024post,smite_new}, the hybrid work context is also changing \cite{smite_new}. Hence, further studies on hybrid work practices and methods could contribute to the future way of working after the pandemic. Many organisations no doubt ask what should happen to this change in working patterns. Should it be finished or continued? Moreover, if so, to what extent? And in what policy should be laid down in organisations? Considering the various open venues, we formulate the following research question (RQ) that this study seeks to tackle: 
\textbf{\textit{RQ: What are the software companies' responses to the hybrid working?}}\\
We draw upon the empirical data from a large-scale survey of worldwide software companies to answer the research question. The survey was conducted and translated into seven languages. We included responses from more than 100 software companies in the data analysis. The main result of our study is a taxonomy of software companies' responses to hybrid working. These responses are classified under the three levels (\textit{individual, team and organisation}) of framework that incorporates multiple perspectives on hybrid work in software engineering \cite{paas}. We found these responses witnessed either a positive or negative impact of hybrid working in software companies. Geographical and time ease, better work-life balance, productivity and efficiency were some of the most perceived positive responses. Less productivity, more investment and pressure were some of the most perceived negative responses.

The rest of the paper is structured as follows. In Section \ref{brw}, we mention the background work, followed by a discussion on the research methodology undertaken in Section \ref{sec:research_approach}. Next, Section \ref{find} presents the findings, and Section \ref{sec:discussion} discusses these results from the existing literature. Finally, the conclusion and future work are covered in Section \ref{sec:conclusion}.

\section{Background and Related work} \label{brw}
\label{brw}
Research interest in hybrid work has surged across disciplines since the first quarter of 2020 \cite{khanna2024hybrid}. Working outside the premises of the offices remotely, from home or from anywhere is defined as a flexible or hybrid way of working \cite{kemell2023hybrid}. This way of working is repeated intervals of work carried out of the office as a practice one or more times per week that provides freedom and openness to the organisation employees regarding working hours and rescue home-office-home commute time \cite{hil}. In the context of software organisation, hybrid software development is characterised by team members working from anywhere (i.e. home or office) and intermittently touching base with the office \cite{conboy2023future}. Literature associated with the WFH practice has mentioned the positive impact of increased productivity \cite{butler} and cost savings \cite{ruth}. One study remarks that, in 2004, it was difficult to find studies that do not witness that telecommuting does not create productivity gains\cite{west}. A global, longitudinal survey study \cite{russo} reports similar findings. Despite arguing that WFH decreased productivity during the pandemic, the authors found that developers are more focused when working remotely, and it does not affect task completion. The well-being and joy of software developers play an important role in the quality of work output, motivation, and henceforth, the success of the entire software development cycle \cite{turkey}. 

A remote work study describes the use of online tools with call centre personnel, e.g., in \cite{bloom}, then geographically distributed software development (GSD) using digital tools to collaborate \cite{rama_gdsd}. Going into individual studies, the study by Neumann et al. \cite{neumann}, for example, reports that the performance of the agile teams in the case organisations of their multiple case studies did not decrease due to remote work. The team members felt they improved their way of working as the online tools made their agile approach more transparent \cite{khanna2022know} and the work process more effective \cite{neumann}.  Additional benefits reported in the study included less online time than offline meetings. In the context of a single software organisation, the study by Canna et al. \cite{canna} reports productivity gains. The study finds that productivity gradually increases during the pandemic as developers adjust to the novel WFH context. This way of working benefits organisations as it helps retain and attract employees. Further, it gives rise to proving the work commitment at the organisational level. Moreover, attracting new employees with the flexibility of working hours \cite{aboe}. Grey literature occasionally argues adverse effects, claiming it could reduce productivity instead of increasing it. However, it is not suited for everyone. The studies reported that some individuals who initially expressed interest in WFH in their case company eventually concluded that it was not for them, choosing instead to return to the office of their own volition. These individuals were also some of the poorer WFH performers \cite{bloom}. Indeed, some individuals may thrive in a more autonomous work environment, while others may have trouble staying disciplined on their own\cite{bloom}. Studies have also discussed challenges with work-life balance \cite{silv} and increased workload \cite{smite} as challenges associated with the pandemic WFH situation. In terms of well-being, the pandemic might also disproportionately have affected women, parents \& people with disabilities \cite{ralph_pp}. 
    
The study by \cite{rama_gdsd} found that remote work distribution could reduce software project team performance due to coordination and administration problems, even with the correct use of enabling technologies. The study by \cite{baker} argued that the findings of different WFH studies are hardly comparable due to the narrow scope of the studies or how they only focused on an individual variable. Moreover, based on existing studies, the main challenges would be a lack of social contact \cite{miller}, feelings of isolation \cite{silv}, and informal social conversations \cite{mach}. To specific factors potentially affecting well-being, Canna et al.\cite{canna} discuss potential issues related to sharing work with childcare at home, unsuitable home environments for remote work, loss of contact with higher-ups, and other reported concerns such as financial issues related to the pandemic situation. Other factors include sleep quality, exercise, decision latitude, work-life balance \cite{silv}, and job strain \cite{turkey}. Numerous studies argue that the pandemic situation has impacted software development \cite{russo,ralph_pp,canna,turkey}. Some studies reported decreased productivity \cite{ralph_pp,russo}, while other studies found no changes in productivity \cite{smite,silv,neumann}, or even increased productivity \cite{canna,bez_sdt}. This study aims to complement the existing studies by investigating the trend of software companies adopting WFH after the pandemic.

\section{Research Approach}
\label{sec:research_approach}
\begin{figure}
\centering
\includegraphics[width=\columnwidth]{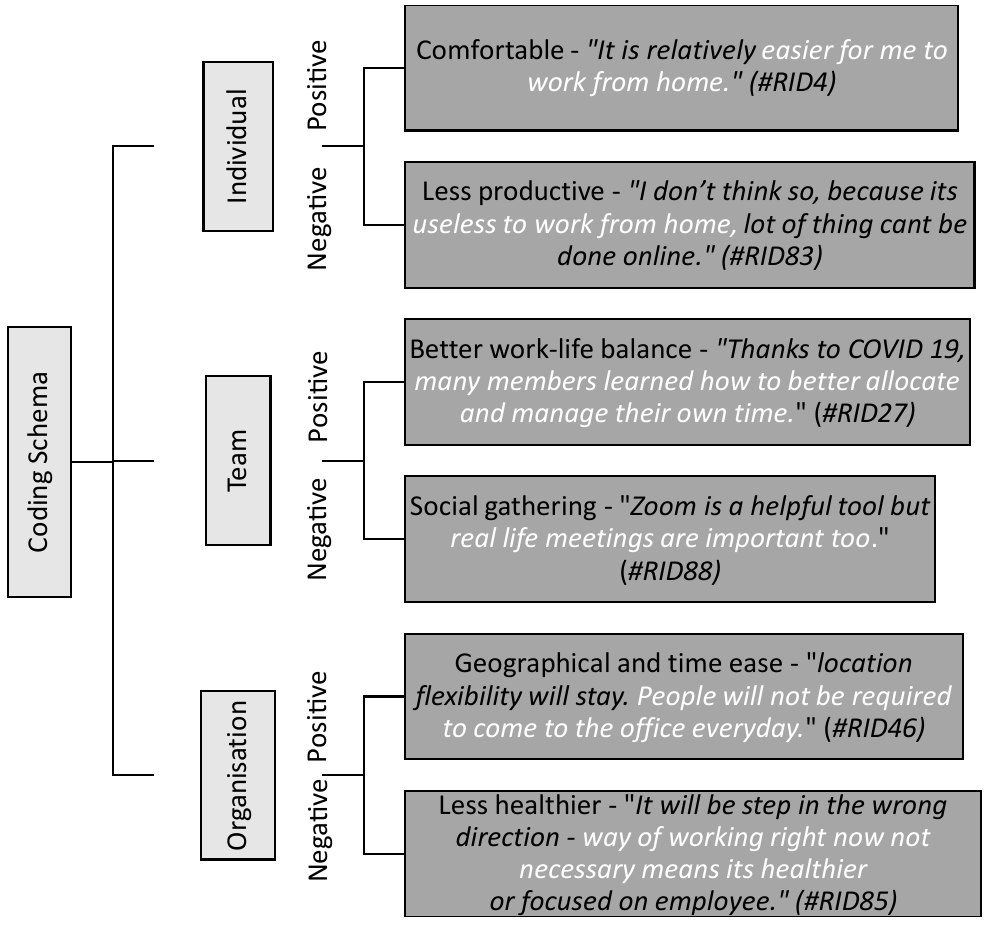}
\caption{\textbf{Coding schema}}
\label{fig:neut}
\end{figure}
This study is based on a large survey \footnote{The survey link can be found \textcolor{blue}{\href{https://covidnse.limesurvey.net/561361?newtest=\&lang=en} {here.}}} that aimed to understand the impact of the COVID-19 pandemic and the hybrid work environment on software companies. For this study, we only used a subset of the questions in the survey. These questions are composed of three parts. The first part is the screening question: \textit{``Do you experience or observe an impact of COVID-19 on your work/company to any extent?"}. If the answer is no, the respondent is navigated to the end of the survey. It ensures validity and only gathers answers from the respondents who experienced the change in the working environment. In the second part, the respondents were asked to provide background information about their roles and companies, including the countries, the team or company size, and business sectors. In the third part, the participants were asked the questions: \textit{``What do you think about new \& effective ways of working emerge due to COVID-19 that will remain after COVID-19? Has your company discovered new business opportunities that have emerged due to COVID-19? If yes, please elaborate. Can you elaborate on some of your answers to the factors contributing to your company's resilience?''} 

\subsection{Data collection}
The survey was developed and tested in three iterations before rolling it out globally. The first version obtained five responses, the second version 25, and the third version had five answers. We created the first version of the questionnaire, based on the software engineering literature and the COVID-19 pandemic, using Google Forms. After rolling it out for around one month, we created the second version based on the collected responses. Also, we considered the comments from the authors to improve the survey questions. We re-coded labels of some questions, removed a few questions for precise focus, added the team or company size and age of the software company, and added applicable questions. Finally, the third version added more open-text questions for the qualitative data analysis. The third version of the survey was made available online via the Lime survey tool. We conducted 12 meetings to reach the last version of the survey study from April to August 2021. The questions were translated into seven languages, Italian, Spanish, Portuguese, Norwegian, Arabic, Indonesian, and Vietnamese, with various native experts. The survey consisted of several demographic questions, and the completion time ranged from 4 to 155 minutes. 

In total, 324 complete responses were received.
Then, we filtered out incomplete responses and excluded responses if the answers were redundant or repeated for all questions. As a result, the total sample size was reduced to 124 \footnote{The 124 responses can be found \textcolor{blue}{\href{https://docs.google.com/spreadsheets/d/1s6hHpLYrIuFcmldVZr2bbiYEaZVVE_Yd8on7rwWT13Y/edit?usp=sharing} {here.}}}. 

\subsection{Data analysis}\label{da}
Paasivaara and Wang \cite{paas} introduced a framework for organizing research on hybrid work in software engineering into three different categories in terms of people's perspectives.
The \textit{people} perspective is divided into \textit{individual, team,} and \textit{organisational} levels. At the individual level, personal factors influence the choice of work mode or way of working. At the team level, the work arrangements influence where individuals work either partially, fully onsite or remotely. Smite et al. \cite{smite199} also propose utilising several team categorisations with the spectrum of hybrid work arrangements. At the organisational level, companies' factors and policies influence their choice of work mode or way of working. We applied thematic analysis, commonly seen in empirical SE research \cite{cruz}.
First, we familiarised ourselves with the data. We read all the responses to evaluate the quality and identify possible themes. We looked for texts that helped us to answer the research question. Text that is irrelevant or gives no helpful content was excluded. Then, we revised the themes, and after validating and taking feedback, we generated themes or higher codes. Later, we classified our higher codes based on the three \textit{individual, team,} and \textit{organisational} levels as used in \cite{paas}. The code taxonomy is shown in Figure \ref{fig:neut}.
Each response was further sub-classified into positive or negative responses, as shown in Figure \ref{fig:neut}. Positive responses witnessed a \textit{positive impact on the three levels}, and negative responses signify \textit{negative impact on the three levels}. For example, reading from top to bottom and right to left, in Figure \ref{fig:neut}, the fourth respondent mentioned \textit{``It is relatively easier for me to work from home.''}(\#RID4). This shows the positive impact of hybrid work on the individual. Hence, the response was categorised under the higher code \textit{comfortable} and \textit{individual} level.

\section{Findings} 
\subsection{Demographic of Respondents}
\begin{figure}[H]
\begin{tikzpicture}
\pie[
        /tikz/every pin/.style={align=center}, 
        text=pin,
        polar,color={gray!10,gray!20,gray!30,gray!40,gray!50,gray!60,gray!70,gray!80,gray!90}]{19/Brazil, 18/UK, 10/Portugal, 10/Italy, 7/Poland, 6/USA, 5/Greece, 3/\\\\Vietnam, 46/Others}
 \end{tikzpicture}
\caption{\textbf{124 survey respondents from 24 countries}}
\label{fig:countries}
\end{figure}
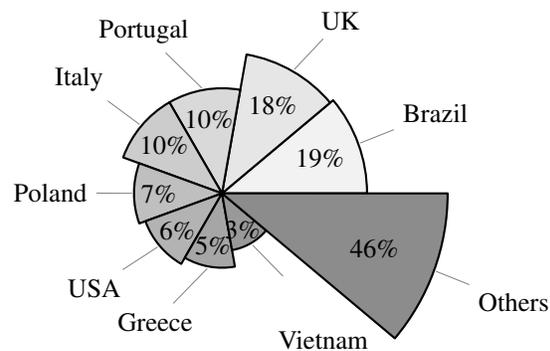

Following is the demographic information of 124 respondents. As shown in Figure \ref{fig:countries}, the responses came from 24 countries and were dominated by respondents from Brazil, the UK, Portugal, and Italy. 
\begin{table}
\caption{Software companies working sector}
\label{tab:worksec}
\centering
\footnotesize
\begin{tabular}{p{.2cm} p{5.2cm} p{1.7cm} } 
  \hline
 \textbf{No.} & \textbf{Work sector} & \textbf{\# of Respondents} \\ \hline 
1	& \textit{Information \& Communication}  & 25 \\
2 & \textit{Banking, Insurance \& Finance}	& 21\\
3 & \textit{Professional, scientific, technical activities (software consultant, auditing, marketing, research, etc)}	& 15\\
4 & \textit{Education}	& 11\\
5 & \textit{Automotive (autonomous car, etc)}	& 8\\
6 & \textit{Healthcare}	& 7\\
7 & \textit{Entertainment \& Games}	& 5\\
8 & \textit{Wholesale \& Retail Trade}	& 5\\
9 & \textit{Energy (green, sustainable energy, etc)}	& 3\\
10 & \textit{Chemical}	& 2\\
11 & \textit{Other}	& 22\\ \hline
 & TOTAL RESPONDENTS	& 124\\ \hline
\end{tabular}
\end{table}
\begin{figure}[H]
\begin{tikzpicture}
\pie[
        /tikz/every pin/.style={align=center}, 
        text=pin,
        polar,
        color={gray!10,gray!20,gray!30,gray!40,gray!50,gray!60,gray!70,gray!80,gray!90}
    ]
    {
        40/Software\\developer, 
        23/Project lead \\ or manager, 
        15/Founders, 
        10/Software architect, 
        10/Business analyst, 
        9/Tester, 
        7/Engineer, 
        5/UX Designer, 
        5/Scrum master 
    }
 \end{tikzpicture}
\caption{\textbf{Roles of 124 survey respondents}}
\label{fig:fig23}
\end{figure}
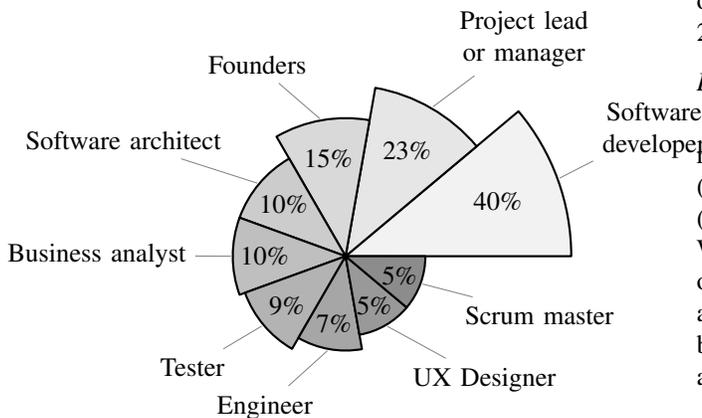
\begin{figure}[H]
\begin{tikzpicture}
\pie[
        /tikz/every pin/.style={align=center}, 
        text=pin,
        polar,color={gray!10,gray!30,gray!50,gray!60,gray!80}]{17/(2-4) \\people, 32/ (5-9) people, 41/(10-49) people, 21/(50-249) people, 13/250\\ or more \\people}
 \end{tikzpicture}
\caption{\textbf{124 survey respondents working in different team or company size}}
\label{fig:66}
\end{figure}
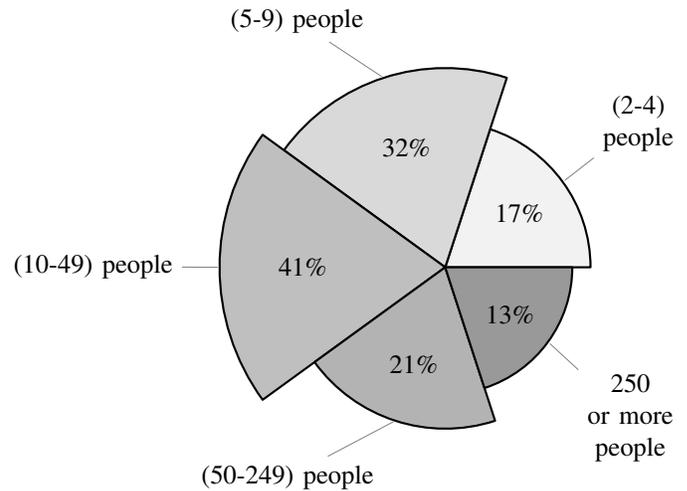

Others specify the following countries (India, China, Finland, Australia, Denmark, Latvia, Japan, Syria, Ireland, Mexico, South Africa, Slovenia, and Sweden). Sixty-six per cent of respondents claimed to work in established companies. In contrast, 22\% respondents worked in a startup, while 12\% were unsure which sector the company belongs to. Sixty-two per cent of respondents said that COVID-19 has impacted the way of work in the companies, whilst 19\% were unaffected by COVID-19. Respondents worked in software companies with different year spans that existed from 1 to more than 10 years. Table \ref{tab:worksec} presents the software companies' working sector. Most companies provide software products or services in the information and communication sector (25 respondents), followed by the banking, insurance and finance sectors (21 respondents). 
Figure \ref{fig:fig23} shows the respondent's role in the company, such as software developers, project lead or managers, business analysts, testers, CEOs or CTOs, engineers, etc. The respondents worked in different teams or company sizes, as shown in Figure \ref{fig:66}. Respondents belong to a minimum team or company size of 2-4 people and a maximum of more than 250 people.
\subsection{Companies Responses to the Hybrid Working}
\label{find}
This section presents an overview of the findings of our research question. We identified 16 participant response codes (see Table \ref{tab:positive_responses}). Most responses positively impacted hybrid work (64\%), while only (36\%) perceived hybrid work negatively. We did not find a significant difference in terms of perception of the impact of the hybrid way of working between startups and established companies. However, there was less difference between the survey respondents who belonged to a small team and those working with a large team.

\begin{table}
\caption{Summary of positive and negative impact from respondents }
\label{tab:positive_responses}
\begin{tabular}{p{1.5cm} p{0.3cm} p{4cm} p{1.2cm}} 
 \hline 
 \textbf{Response code}  & \textbf{Occ.} & \textbf{Example of quote} & \textbf{Level} \\ \hline
  & &  \textbf{POSITIVE IMPACT} &\\ 
 \hline 
        Better work-life balance & 13\%  & \textit{``Thanks to COVID-19, many members learned how to better allocate and manage their own time. ''}  & Team\\
        
        Comfortable& 12\% & \textit{``You know, I really like having meetings and doing most of the work online as it's soothing.''} & Team \\
        
        Ease of online meetings & 4\% & \textit{``Online teams got way better during the pandemic.''} & Team \\
        
        Ease of remote work & 2\% & \textit{``I think that with the pandemic, it has been proven that it is possible to work remotely with quality, at least in the IT area.''} & Organisation \\
        
        Economic savings& 6\% & \textit{``Certainly, many companies will stay with remote work as more economical, for example renting an office.''} & Organisation \\
        
        Geographical \& time ease & 15\% & \textit{``Working from home will definitively be more frequent in general due to time savings in transportation and prep work.''} & Individual \\
        
        Productivity \& efficiency& 13\%  & \textit{``I think the companies need to implement this new way of working from home; as far as I'm aware, the productivity has increased.''} & Organisation \\
        
        Reliance on online tools & 2\%  & \textit{``Remote working enhanced online connectivity, optimize operational processes through digital tools.''}& Team \\ 
        
        \hline
       & & \textbf{NEGATIVE IMPACT} &\\ 
        \hline

Lack of job satisfaction& 3\% & \textit{``Personally, I've found remote working distracting, demotivating and inefficient; however, it is inevitable that generally will now increase. ''}  & Individual\\

        Lack of skilled people& 2\% & \textit{``Our company had a significant decrease in employees working way in person during the pandemic due to slightly older age.''} & Organisation \\
        
        Lack of work-life balance& 2\% & \textit{``my company to change the way we do this which has further put stress on the company employee.''} & Organisation \\
        
        Less healthy& 2\%  & \textit{``In my opinion it will step in the wrong direction - way of working right now not necessary means its healthier or focused on the employee.''}& Individual\\
        
        Less productive& 9\% & \textit{``I don't think so, because it is useless to work from home, a lot of things can't be done online.''} & Individual\\
        
        More investment & 9\%  & \textit{``We are considering alternative work arrangements; my organisation has to invest in new tools to facilitate remote work like Microsoft Teams and M365.''}& Organisation\\
        
        More pressure & 3\% & \textit{``The COVID-19 situation working has meant we have had to work harder and longer for the same benefits as before.''} & Team \\
        
        Lack of social gathering & 3\% & \textit{``There will always be a need for face-to-face contact and, although excellent and essential, it is not a 100\% substitute for the direct contact which was normal before COVID.''} & Team \\ \hline
\end{tabular}
\end{table}
Table \ref{tab:positive_responses} shows the list of positive response codes, examples of quotes and response code occurrences (occ.) to hybrid work. The most positive responses perceived by the respondents were the geographical and time ease, better work-life balance, and increased productivity and efficiency. One respondent said that \textit{``Everything that can be done remotely will help impact commuting, traffic, time spent in transport and perhaps a long-term breath on the environment.''} (\#RID61). Moreover, another respondent pointed out that working from home gives the possibility of being closer to the family and improving the quality of work-life balance. \textit{``With this extra time and the possibility of being closer to the family, you have a much higher quality of life. I have a small son, and my relationship with him has improved a lot after the pandemic.''}(\#RID21). 

Table \ref{tab:positive_responses} summarises and shows the list of positive response codes, examples of quote and response code occurrence to hybrid work. Among the negative responses, the respondents reported that the shift to hybrid working after the pandemic led to less productivity and more investment needed by the organisation. As one of the respondents admitted, some activities cannot be performed online.\textit{``Employees perform activities that cannot be performed remotely. So much of the innovation is somewhat stalled.''}(\#RID107). 

As mentioned in the previous section \ref{da}, we categorised these 16 response codes under the three levels depending on the impact participants captured with a hybrid of working.

\noindent \textbf{Individual level responses}
\begin{figure}
\centering
\includegraphics[width=\columnwidth]{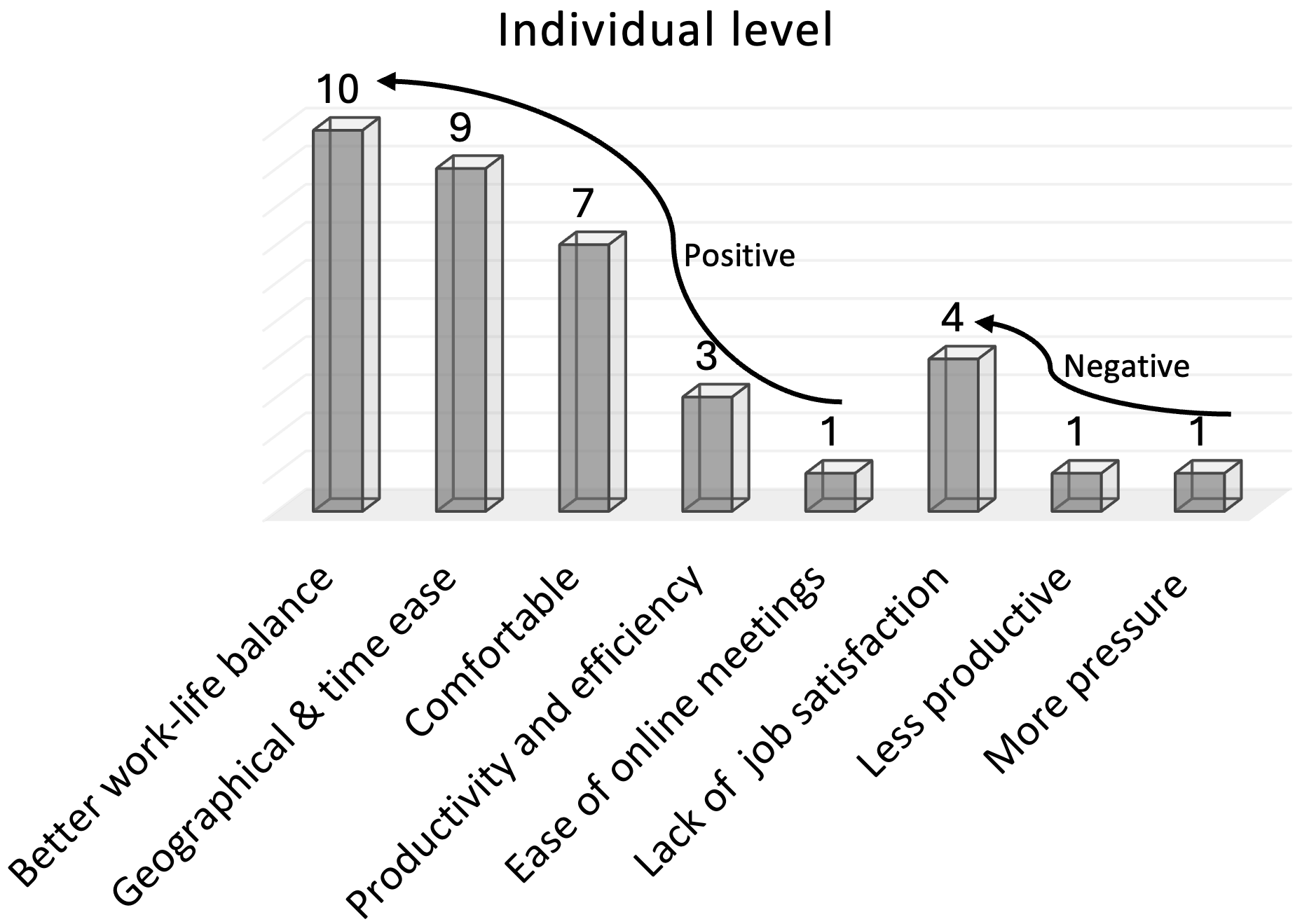}
\caption{\textbf{Responses categorised based on individual level}}
\label{fig:indi}
\end{figure}
At the individual level, evident from Figure \ref{fig:indi}, participants perceived hybrid working positively, i.e. five positive (30 responses) vs. three negative codes (6 responses). Hybrid working is acknowledged as contributing to better work-life balance. A respondent mentioned \textit{``Working from home...actually helps you to better regulate yourself''}(\#RID51). At the same time, respondents also realised the geographical and time ease of working with the hybrid way of working. Respondent mentioned \textit{``Significantly reduce the time wasted on moving, waiting''}(\#RID50). \textit{``Don't have to spend hours to commute''}(\#RID68). Hybrid working was also considered comfortable for individuals as respondents addressed \textit{``effective work can be done from the comfort of your own home''}(\#RID81). Looking at Figure \ref{fig:indi}, we can observe several other response codes which show a positive and negative impact of a hybrid way of working. For example, positive impacts include increased productivity, efficiency, and ease of online meetings. On the other hand, negative impacts include a lack of job satisfaction, less productivity, and more work pressure. 

\begin{figure}
\centering
\includegraphics[width=\columnwidth]{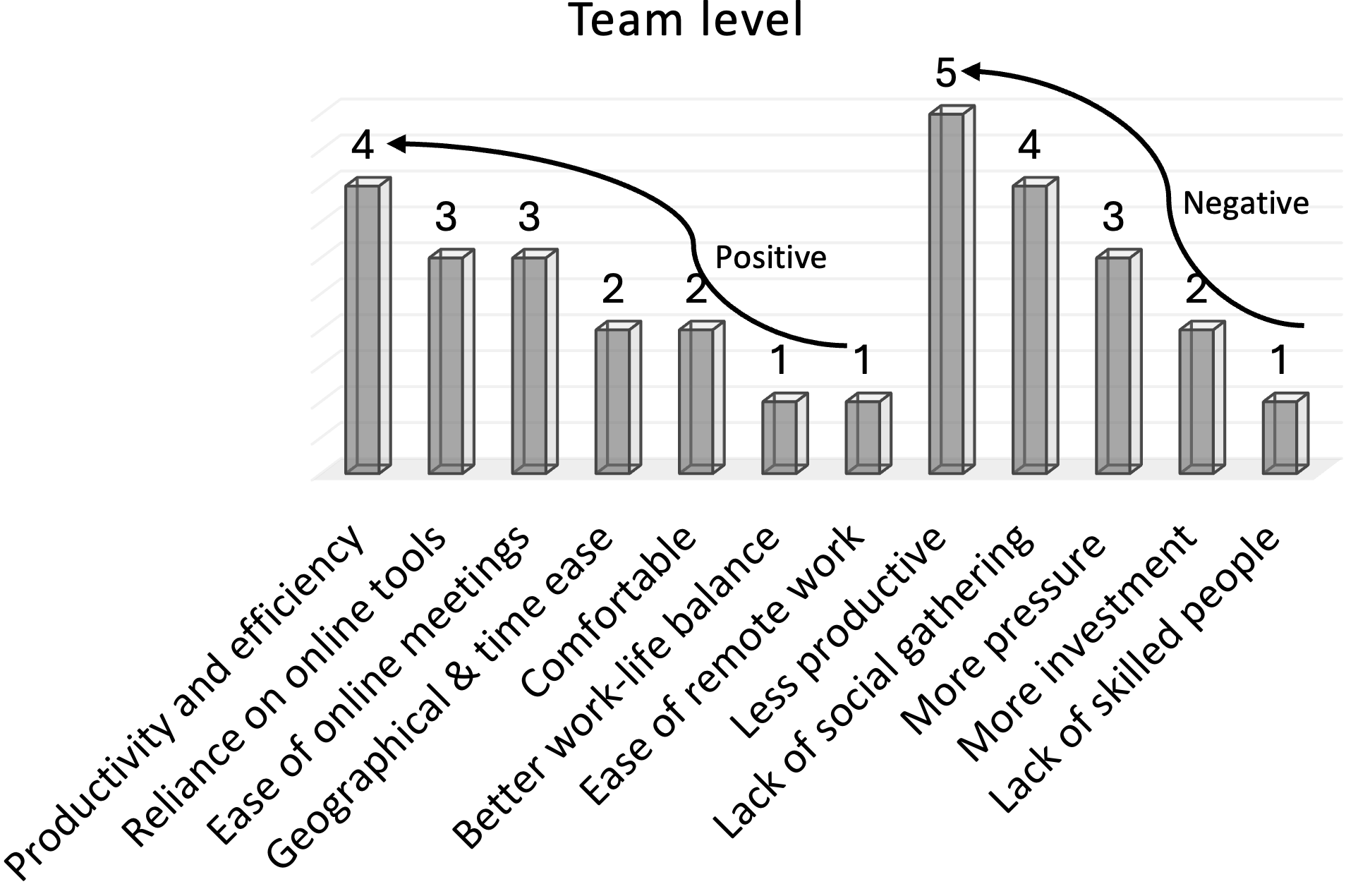}
\caption{\textbf{Responses categorised based on team level}}
\label{fig:team}
\end{figure}

\noindent\textbf{Team level responses} At the team level, the positive and negative impacts of hybrid working were equally mentioned by respondents. We identified 16 occurrences of positive, compared to 15 negative responses. 
Evident from Figure \ref{fig:team}, participants acknowledged less productive the hybrid way of working. A respondent mentioned \textit{``We currently fail as working in teams at the moment''}(\#RID105). At the same time, respondents also realized the need for social gathering is crucial, which is difficult with hybrid working. Respondent mentioned \textit{``When you aren't face to face to communicate, it is quite hard to follow direct orders or communicate''}(\#RID108). Hybrid working was also seen as productive and efficient for teams as respondents addressed \textit{``The coordination in teamwork is also increased. More proactive in their work, and use more tools to improve work efficiency.''}(\#RID28). Looking at Figure \ref{fig:team}, we can observe several other response codes which show positive and negative impacts of a hybrid way of working. For example, positive impacts, ease of remote work and online meetings, reliance on online tools, etc. Whereas negative impacts, lack of job-skilled people, more investment in teamwork, etc. 

\begin{figure}
\includegraphics[width=\columnwidth]{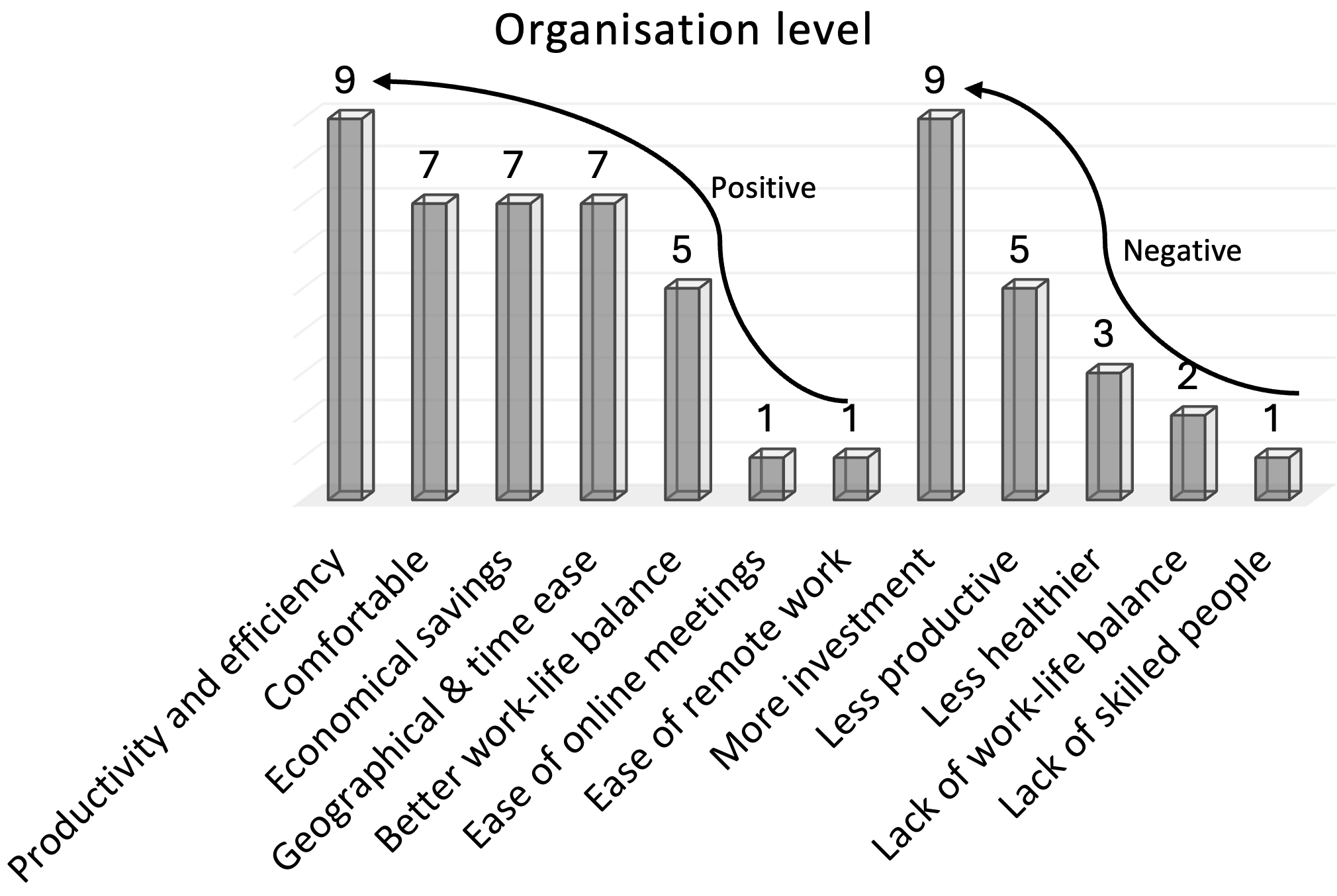}
\caption{\textbf{Responses categorised based on organisation level}}
\label{fig:organisation}
\end{figure}

\noindent\textbf{Organisation level responses}
At the organisational level, we identified 37 occurrences of positive responses and 20 negative responses. As evident from Figure \ref{fig:organisation}, participants mentioned productivity and efficiency as the most repetitive responses. A respondent said \textit{``I think the companies need to implement this new way of working from home,..., the productivity has increased''}(\#RID16). 
At the same time, respondents also witnessed more investment is needed in the hybrid way of working to run the organisation. The respondent said \textit{``We are considering alternative work arrangements; my organisation has to invest in new tools to facilitate remote work like Microsoft Teams and M365''}(\#RID113). \textit{``We had to invest, innovate and change how customers buy a new product that is no longer in the store, but in the APP.''}(\#RID112). At the organisation level, hybrid working was also seen as comfortable as respondents addressed \textit{``Home office should continue, as it is more comfortable for workers''}(\#RID124). Looking at Figure \ref{fig:organisation}, we can observe several other response codes which show positive and negative impacts of a hybrid way of working. For example, economic savings, geographical and time ease, etc., as positive impacts. On the other hand, negative impacts include less health, lack of work-life balance, etc. 


\section{Discussion} \label{sec:discussion}
\begin{figure}
\centering
\includegraphics[scale=0.63]{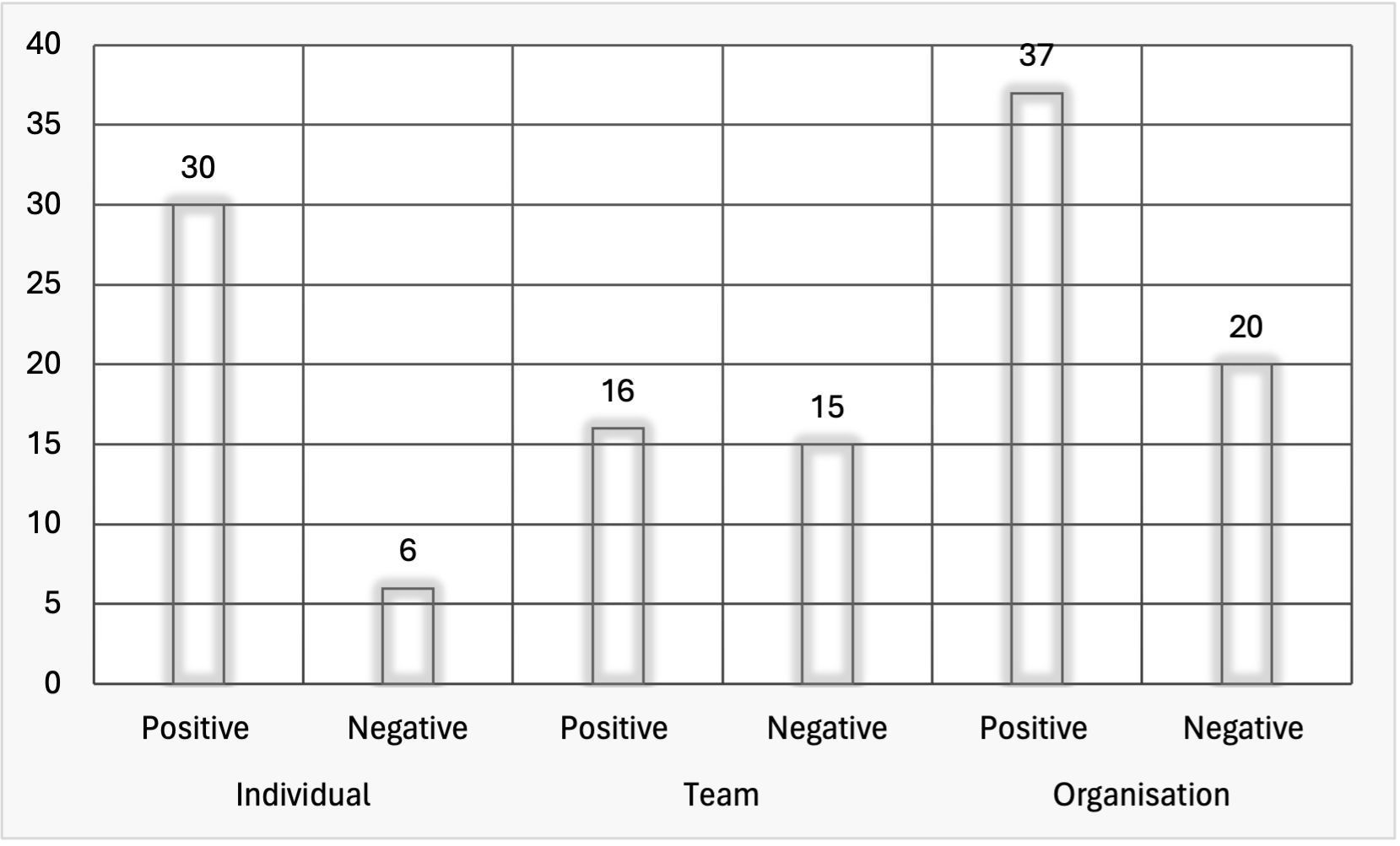}
\caption{\textbf{Comparison of responses at three levels}}
\label{fig:comp}
\end{figure}

We conducted a large-scale survey to answer our research question and obtained 124 responses. Each of the responses was classified under the three levels of people perspective \cite{paas}. Further, each response was sub-classified under the respective level as either positive or negative. A positive response witnessed a positive impact of hybrid working, whilst negative responses certified a negative impact at particular levels. Evident from Figure \ref{fig:comp}, we found a higher positive impact at the individual and organisation levels. At the team level, hybrid working has similar positive and negative impacts. Hence, this leads to a count of the higher positive impact of hybrid working. We observed positive outweigh 83 responses to negative 41 responses.

Considering the 16 response codes shown in Table \ref{tab:positive_responses}, we observed eight positive and negative responses. In some instances, we found some of the response codes with positive and negative impacts, for example, \textit{work-life balance}. For some respondents, hybrid working proved to be positively impacted at the different levels (see Figure \ref{fig:indi}, \ref{fig:team}, and \ref{fig:organisation}), hence denoted as better work-life balance. In comparison, the same response code for some respondents' hybrid working was perceived to be negatively impacted at the organisation levels as shown in Figure \ref{fig:organisation}, hence denoted as a lack of work-life balance.

Hybrid working provides flexibility and a self-government software development process (\#RID57). Developing software does not mandate the physical presence of employees in the workspace. This way of working to develop software at startups or established companies give the freedom to choose the working location depending on individual or team needs (\#RID26). In terms of team collaboration, online tools play a major role. For example, using Slack, Jira, Trello, Zoom, Microsoft Teams, etc., for communicating and collaborating on a project (\#RID28). Using the online tools \cite{khanna2022your} in meetings \cite{khanna2023exploring} focuses the effort of an individual on working and delivering the software, rather than time and effort spent (\#RID10) while commuting to the office \cite{nguyen2023work}. This pattern of working and developing the software encourages a better work-life balance. Hence, allowing more time with their family members and individuals can pursue personal interests and hobbies (\#RID21). 

In comparison to the literature (see Section \ref{brw}), we found similar themes that show hybrid work increases productivity \cite{butler}, and it is a cost-saving practice \cite{ruth}. A company can cut unnecessary costs like travel and physical space. Another positive impact of hybrid work is employee retention. The employee prefers to be retained longer for the organisation. If employees could coordinate, trust, unite, and have a strong bond in a team.  Many organisations already were taking advantage of digital technologies \cite{newford}, which made hybrid work \cite{nolan} a future way of working with increased productivity \cite{bailey}, but for newly established firms or startups with a shortage of resources \cite{paper1}, it was challenging. Whereas, from \ref{brw}, for some respondents, hybrid working decreases productivity \cite{bloom}, as the employees are demotivated and inefficient at work as they cannot maintain coordination \cite{rama_gdsd}, and several things cannot be done online. Firms should invest more in social gatherings to keep the employees connected \cite{cuco}. However, we cannot ensure the same rule applies to all organisations. Staying focused, motivated, and not getting distracted is a challenge \cite{russo_21}. Our respondents also validate this; certain employees found it a negative way of working. Employees demanded \textit{``It really affects our scheduled meetings and the effectiveness of our meetings''}(\#RID118).

Regarding threats to the study's validity, a threat to our study is the association between a study and its outcome \cite{wohlin}. The survey contained the screening question to overcome the threat: Most respondents (61.70\%-Yes, 19.14 \%-No, 13.19\%-No Answer, 5.95\%-Don't know) answered that COVID-19 impacted the work. A threat is that respondents in the survey can only express their perceptions. To overcome this, we revised the survey design in several iterations with researchers at the Software Startups Research Network. The construct validity explains the connection between the investing theory and the observation conducted with the survey \cite{wohlin}. This study aimed to obtain the company's insights regarding WFH practices and their effects. To overcome this, we used authentic software engineering scales and were confident about the responses, reducing biases in the answers.

Regarding the limitation of our study, a complete lockdown could have influenced the respondent’s attitudes to certain effects differently. Participants might be biased with the answers as they could have seen digital tools’ communication pattern or importance differently if it was not a complete lockdown. The restriction has been lifted in many countries; thus, future studies could investigate and examine our findings.

It is further strengthened with the help of data triangulation, which is having a consistent observation of the data. We did our best not to conduct random respondents. We categorized them based on their geographical location, industrial sections, and company types. Another limitation is that many of the responses were written in the same language as the survey. As we did not have expertise in all survey languages, we relied on online translation tools to understand and analyse the responses in English. There might be some missing interpretations of the responses.

\section{Conclusion} \label{sec:conclusion}
The COVID-19 pandemic has caused various adverse effects worldwide. Many work sectors were halted or disrupted due to the pandemic. The software industry witnessed the impact of COVID-19 and the hybrid way of working. Our results are from a qualitative analysis of 124 survey responses. We applied thematic analysis to code the qualitative data. We found positive and negative impacts of the hybrid way of working. The change in the work pattern shows respondents in organisations were more united and better aware of work goals. There was a noticeable increase in productivity for some software companies (established companies or startups). Things were available online due to substantial digital transformation and changes in company dynamics. Teams became more decisive and responsible in terms of the workflow. With these mentioned positive impacts, there were negative impacts, too. For example, a few individuals mentioned remote work is efficient, but they felt distracted, demotivated, and stressed as things could not be done online. The team felt the need for face-to-face contact and essential physical office time.

\section{Acknowledgments}
We would like to express our appreciation to those \href{https://softwarestartup.org/about/}{SSRN} participants who played a major role in the survey design and collection of responses. This work has been supported by ELLIIT; the Swedish Strategic Research Area in IT and Mobile Communications.

\bibliographystyle{IEEEtran}
\bibliography{sample-base}


\end{document}